\documentclass[%
    reprint,
    superscriptaddress,
    amsmath,amssymb,
    aps,
    prl,
    floatfix,
    longbibliography
]{revtex4-2}

\usepackage{graphicx}
\usepackage{xcolor}
\definecolor{revisiongreen}{RGB}{0,128,0}
\definecolor{revisionyellow}{RGB}{153,112,0}
\usepackage[percent]{overpic}

\usepackage{dcolumn}
\usepackage{bm}
\usepackage{array}
\usepackage[colorlinks=true,linkcolor=blue,citecolor=blue,urlcolor=blue]{hyperref}

\newcommand{\figref}[1]{\hyperref[#1]{Fig.~\ref*{#1}}}
\newcommand{\tabref}[1]{\hyperref[#1]{Table~\ref*{#1}}}
\newcommand{\eqnref}[1]{\hyperref[#1]{Eq.~(\ref*{#1})}}

\begin{document}


\title{Mixed-Sign Kitaev Exchange in Jahn-Teller-Distorted CrI$_2$}

\author{Shiyan Gao}
\altaffiliation{Contributed equally to this work.}
\affiliation{Key Laboratory of Computational Physical Sciences (Ministry of Education), Institute of Computational Physical Sciences, State Key Laboratory of Surface Physics, and Department of Physics, Fudan University, Shanghai 200433, China.}

\author{Zekuo Yang}
\altaffiliation{Contributed equally to this work.}
\affiliation{Key Laboratory of Computational Physical Sciences (Ministry of Education), Institute of Computational Physical Sciences, State Key Laboratory of Surface Physics, and Department of Physics, Fudan University, Shanghai 200433, China.}

\author{Chuanhang Chen}
\altaffiliation{Contributed equally to this work.}
\affiliation{Key Laboratory of Computational Physical Sciences (Ministry of Education), Institute of Computational Physical Sciences, State Key Laboratory of Surface Physics, and Department of Physics, Fudan University, Shanghai 200433, China.}

\author{Zefeng Chen}
\affiliation{Key Laboratory of Computational Physical Sciences (Ministry of Education), Institute of Computational Physical Sciences, State Key Laboratory of Surface Physics, and Department of Physics, Fudan University, Shanghai 200433, China.}

\author{Lianchuang Li}
\affiliation{Key Laboratory of Computational Physical Sciences (Ministry of Education), Institute of Computational Physical Sciences, State Key Laboratory of Surface Physics, and Department of Physics, Fudan University, Shanghai 200433, China.}

\author{Hongjun Xiang}
\email{hxiang@fudan.edu.cn}
\affiliation{Key Laboratory of Computational Physical Sciences (Ministry of Education), Institute of Computational Physical Sciences, State Key Laboratory of Surface Physics, and Department of Physics, Fudan University, Shanghai 200433, China.}

\author{Changsong Xu}
\email{csxu@fudan.edu.cn}
\affiliation{Key Laboratory of Computational Physical Sciences (Ministry of Education), Institute of Computational Physical Sciences, State Key Laboratory of Surface Physics, and Department of Physics, Fudan University, Shanghai 200433, China.}
\affiliation{Hefei National Laboratory, Hefei 230088, China}


\begin{abstract}
Kitaev couplings with ferromagnetic (FM) or antiferromagnetic (AFM) character shape magnetic frustration and phases differently in real materials, yet their coexistence within a single material remains unexplored. Here, we propose Jahn--Teller distortion as a route to realizing such coexistence on inequivalent bonds. First-principles calculations identify monolayer CrI$_2$ as such a system, with AFM and FM Kitaev couplings on the short and long bonds, respectively. These opposite signs are robust against variations in Hubbard $U$, functionals, and strain. A minimal tight-binding model links this sign contrast to Jahn--Teller-split $t_{2g}$ orbitals and their hopping to $e_g$ orbitals. 
We further adopt the cluster expansion method and develop a realistic machine-learned spin Hamiltonian, which reproduces the experimentally observed proper-screw geometry and propagation direction and reveals highly nonuniform spin rotation, with turning angles ranging from $48.0^\circ$ to $93.9^\circ$. Both proper-screw anisotropy and strong anharmonic spin rotation arise from the mixed-sign Kitaev interaction. 
These findings not only establish Jahn--Teller distortion as a microscopic route to mixed-sign Kitaev exchange, but also open a new avenue for exploring novel magnetic phases.
\end{abstract}


\maketitle

\begin{figure}[t]
  \centering
  \includegraphics[width=\columnwidth]{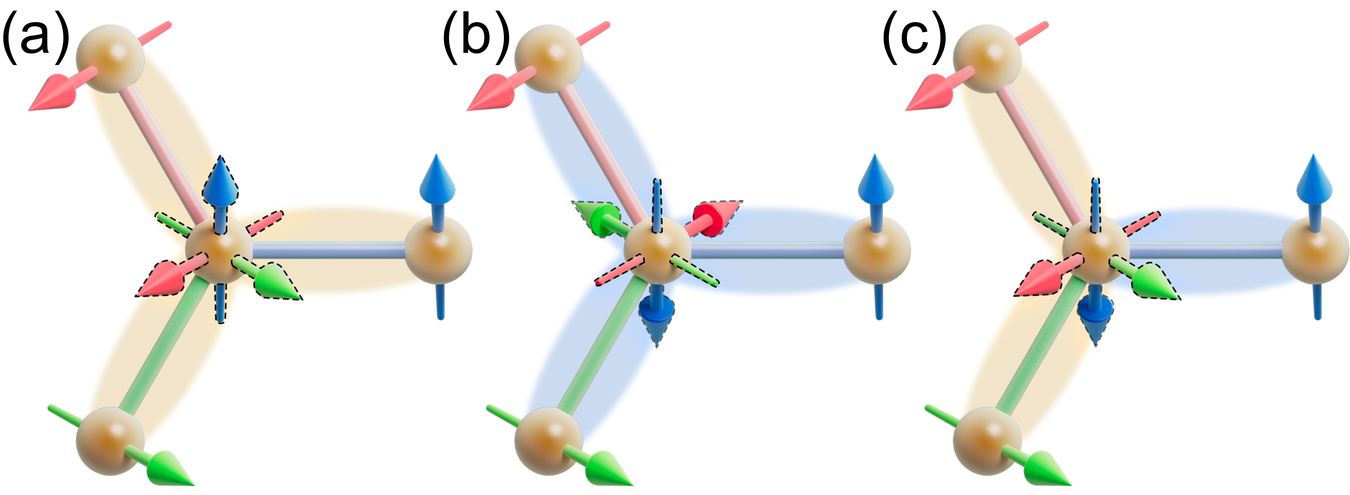}
  \caption{Illustration of frustration induced by differently signed Kitaev interactions. Panels (a) and (b) show the FM ($K<0$) and AFM ($K>0$) cases, respectively, while panel (c) shows the mixed-sign case established here. With the same surrounding spin configuration, the central spin experiences different competing directional preferences.
}
  \label{fig:cri-structure-experiment}
\end{figure}

The Kitaev model provides a paradigmatic route to quantum spin liquids~\cite{kitaev_anyons_2006,hermanns_physics_2018}. On an individual bond, ferromagnetic (FM) $K<0$ favors spins aligned in parallel [\figref{fig:cri-structure-experiment}(a)], whereas antiferromagnetic (AFM) $K>0$ favors antiparallel ones [\figref{fig:cri-structure-experiment}(b)]. In the ideal zero-field model, reversing the signs of the Kitaev couplings, which can be absorbed by spin or gauge transformations, leads to equivalent ground states or Majorana spectra~\cite{kitaev_anyons_2006}. This equivalence, however, is fragile in realistic magnets. Additional Heisenberg interactions already distinguish different signs through their competition with $K$~\cite{chaloupka_kitaev_heisenberg_2010,chaloupka_zigzag_2013,Gotfryd2017,Zhang2021}, while a uniform magnetic field leads to distinct excitation behaviors~\cite{Zhu2018,hickey_emergence_2019,Kaib2019,berke_field_2020}. The sign of $K$ therefore becomes a physically consequential degree of freedom once Kitaev exchange is embedded in a realistic spin Hamiltonian.

Both FM and AFM Kitaev couplings have now been identified in real materials. The canonical $d^5$ edge-sharing mechanism generally favors FM $K$, as exemplified by $\mathrm{Na_2IrO_3}$~\cite{jackeli_mott_2009} and $\alpha$-$\mathrm{RuCl_3}$~\cite{plumb_spinorbit_2014}. In contrast, AFM $K$ has been found in several $3d$ magnets, including $\mathrm{CrI_3}$, $\mathrm{CrGeTe_3}$, and $\mathrm{NiI_2}$~\cite{xu_interplay_2018,amoroso_spontaneous_2020,li_realistic_2023,stavropoulos_microscopic_2019}, with recent experiments further supporting sizable Kitaev interactions in $\mathrm{NiI_2}$~\cite{wang_microscopic_2026,xu_sizable_2026}. Related Kitaev interactions have also been reported in the spin-$1$ magnets $\mathrm{VI_3}$ and $\mathrm{VBr_3}$~\cite{gu_signatures_2024,kao_kitaev_2026}. Polar structural asymmetry has further been proposed as a route to reversing the sign of $K$ by rebalancing competing exchange channels~\cite{sugita_antiferromagnetic_2020}.

As discussed above, the sign-dependent interplay between Kitaev and other interactions can reshape magnetic phase diagrams~\cite{chaloupka_kitaev_heisenberg_2010,chaloupka_zigzag_2013} and even select the propagation direction of helical order, as demonstrated for $\mathrm{NiI_2}$~\cite{li_effects_2024,li_realistic_2023}. A central question nevertheless remains unanswered: can a single material host FM and AFM Kitaev couplings on inequivalent bonds? On a honeycomb lattice, for example, each spin could be coupled to its three neighbors through one FM and two AFM Kitaev bonds, or vice versa [\figref{fig:cri-structure-experiment}(c)]. 
Such a mixed-sign arrangement would favor FM correlations in some spin components and AFM correlations in others. 
Opposite-sign Kitaev couplings were proposed to possibly explain the measured magnetic structure of $\alpha$-Li$_2$IrO$_3$~\cite{Williams2016}, whereas subsequent calculations support all ferromagnetic Kitaev couplings~\cite{Winter2016,Shen2022}.
This motivates the search for materials with robust mixed-sign Kitaev couplings, whose interplay with non-Kitaev interactions may stabilize unconventional magnetic states distinct from those in either uniform-sign limit.

In this Letter, we establish Jahn--Teller distortion as a route to mixed-sign Kitaev exchange and demonstrate it in monolayer $\mathrm{CrI_2}$, by combining first-principles calculations, tight-binding analysis, and a realistic machine-learned spin Hamiltonian. First-principles calculations reveal AFM and FM Kitaev couplings on the short and long bonds, respectively, while the tight-binding model identifies the Jahn--Teller-induced $t_{2g}$ splitting as the key microscopic origin of their sign contrast. The realistic spin Hamiltonian reproduces the experimentally observed proper-screw geometry and propagation direction and further shows that its specific anisotropy, as well as an anharmonic spin rotation, critically relies on the mixed signs of the Kitaev couplings. Finally, phase diagrams obtained under biaxial strain and by tuning the Kitaev couplings demonstrate that their signs play a decisive role in magnetic phase selection.

\begin{figure}[!t]
  \centering
  \includegraphics[width=8cm]{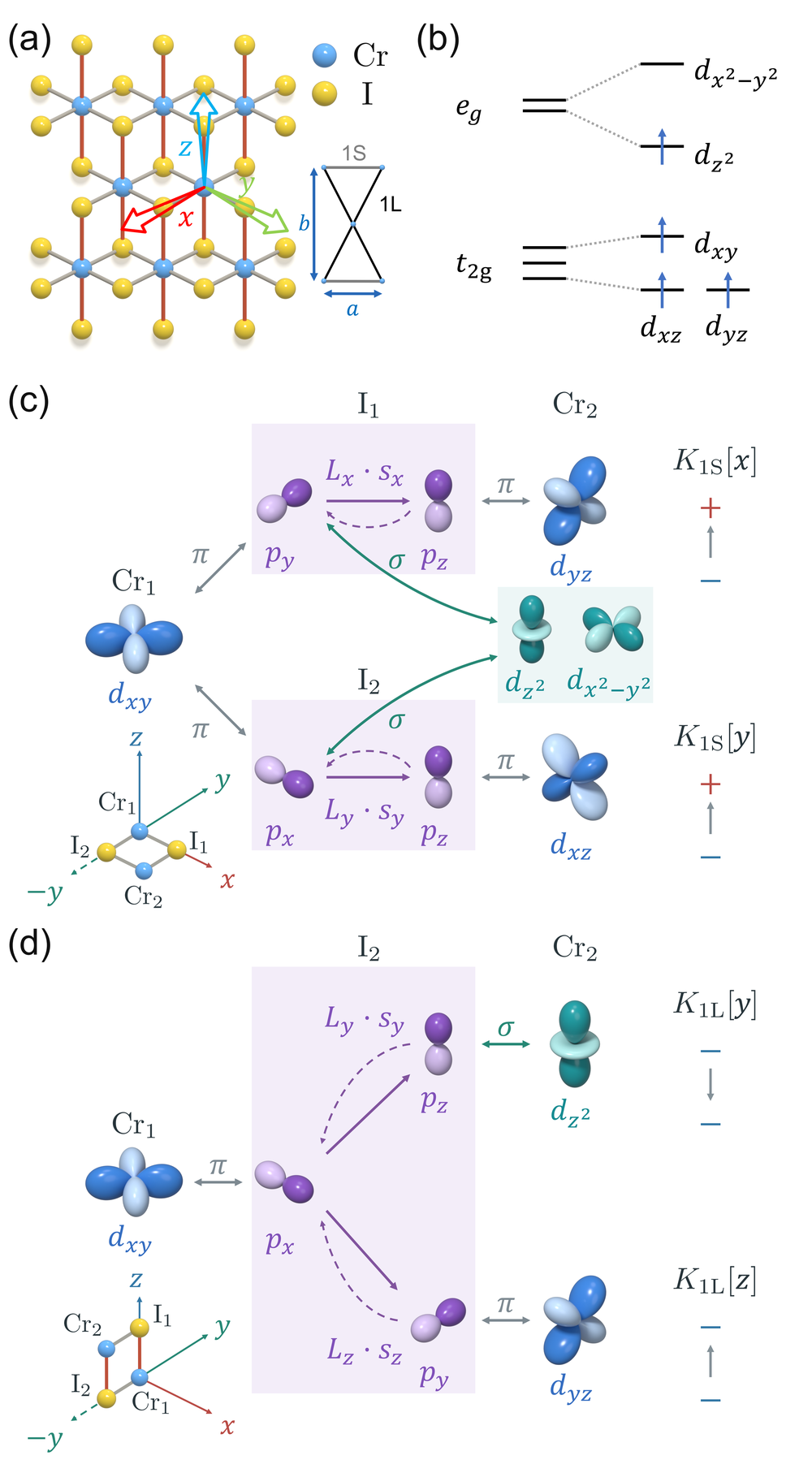}
  \caption{Origin of mixed-sign Kitaev exchange. (a) Crystal structure of CrI$_2$ with gray and brown colors denoting normal and elongated Cr-I bonds, respectively. (b) Jahn-Teller splitting and orbital occupation of high-spin $\mathrm{Cr}^{2+}$ ($3d^4$, $S=2$).  
Panels (c) and (d) illustrate the ligand-SOC-mediated orbital connections that reverse the sign of the Kitaev exchange on the short $1\mathrm{S}$ bonds and preserve it on the long $1\mathrm{L}$ bonds, respectively. Insets show the geometry and orientation of the TB clusters. Purple orbitals are mixed by SOC on the same I site. The displayed connections form part of the coupled $t_{2g}$--$e_g$ hopping network. The signs and arrows beside $K_{\rm 1S,1L}[\alpha]$ show how the contribution from the ligand SOC component $L_\alpha s_\alpha$ changes as $E_{xy}$ increases.}
  \label{fig:guidao}
\end{figure}

\textcolor{blue}{Emergence of opposite-sign Kitaev couplings.}
To obtain markedly different bondwise Kitaev couplings, a natural starting point is to introduce unequal bond lengths, e.g., through Jahn--Teller distortion. Among the Jahn--Teller-active systems considered, we focus on monolayer $\mathrm{CrI_2}$ derived from the monoclinic $C2/m$ bulk structure~\cite{tracy_crystal_1962}. Its $\mathrm{Cr}^{2+}$ ions form a distorted triangular lattice and adopt a high-spin $t_{2g}^{3}e_g^{1}$ configuration with $S=2$ [\figref{fig:guidao}(a),(b)]. Recent experiments have revealed a  proper-screw state in orthorhombic bulk $\mathrm{CrI_2}$~\cite{schneeloch_helimagnetism_2024}. 
The Jahn--Teller distortion lifts the degeneracies within the $e_g$ and $t_{2g}$ manifolds, lowering the energies of orbitals with $z$ character. It produces two elongated axial Cr--I bonds (3.17~\AA) and four shorter equatorial bonds (2.73~\AA). The resulting nearest-neighbor (NN) Cr--Cr network comprises two short bonds ($1\mathrm{S}$, 3.91~\AA) and four long bonds  ($1\mathrm{L}$, 4.23~\AA) [\figref{fig:guidao}(a)].

We construct a first-principles-based spin Hamiltonian to examine the possibility of mixed-sign Kitaev interactions in CrI$_2$. To avoid missing important terms, we adopt the symmetry-adapted spin cluster-expansion method~\cite{lou_pasp_2021}, which systematically generates interactions consistent with the crystal symmetry. We include two-body interactions within the first three coordination shells of the parent undistorted triangular lattice; the structural distortion further splits these shells into inequivalent bonds, which are treated separately in the expansion. We initially include all symmetry-allowed bilinear terms with SOC and biquadratic terms without SOC. A monolayer geometry with experimentally determined lattice constants is adopted~\cite{schneeloch_helimagnetism_2024,sm}. DFT energies of random noncollinear spin configurations are used to fit the model. The resulting spin Hamiltonian reads
\begin{equation}
\label{eq:heff}
\begin{aligned}
\mathcal{H}_{\rm spin}\!=\!
&\sum_{\langle ij\rangle_{1\mu}}\!
\left[
J_{1\mu}\mathbf{S}_i\!\cdot\!\mathbf{S}_j
+K_{1\mu}S_i^{\gamma}S_j^{\gamma}
+B_{1\mu}\left(\mathbf{S}_i\!\cdot\!\mathbf{S}_j\right)^2
\right]\\
&+\sum_{\langle ij\rangle_{3\mu}}\!
J_{3\mu}\mathbf{S}_i\!\cdot\!\mathbf{S}_j
+\sum_i\! A_{n}\left(S_i^{n}\right)^2,
\end{aligned}
\end{equation}
where $\mu= {\rm S, L}$ labels the short and long Cr-Cr bonds within each coordination shell, while $\gamma=x,y,z$ corresponds to the long X, Y, and short Z Kitaev bonds, respectively. The ${\bm n}$ axis of single-ion anisotropy (SIA) lies close to the $z$ direction (see Sec.~I(c) of SM~\cite{sm}).

We first focus on the Kitaev couplings on 1S and 1L bonds.
For classical unit spins, we obtain $K_{1\mathrm{S}}=J_z-(J_x+J_y)/2=+0.48$ meV and $K_{1\mathrm{L}}=J_x-(J_y+J_z)/2=-1.52$ meV. The short and long bonds thus host AFM and FM Kitaev couplings, respectively. Such results are cross-validated using the four-state method~\cite{Xiang2011FourState,Xiang2013EnergyMapping}. Moreover, the opposite signs of $K_{1\mathrm{S}}$ and $K_{1\mathrm{L}}$ persist over a broad range of tested strains, Hubbard $U$ values and  functionals (including HSE), though the values of Heisenberg and Kitaev couplings may vary  [see \textcolor{blue}{Sec.~I(e)} of Supplemental Material (SM)~\cite{sm}]. These results establish mixed-sign Kitaev exchange in Jahn--Teller-distorted CrI$_2$.
We next investigate the microscopic origin of this sign contrast before exploring its consequences for the ground state.

\textcolor{blue}{Microscopic origin of the sign reversal.}
To understand the opposite signs of the Kitaev couplings, we develop a minimal tight-binding (TB) model of a two-Cr--two-I cluster within the single-electron approximation. The short $1\mathrm{S}$  and long $1\mathrm{L}$ bonds are represented by square and rectangular clusters, respectively, with I--Cr--I angles fixed at $90^\circ$ [\figref{fig:guidao}(c),(d)]. The Hamiltonian reads
\begin{equation}
\begin{aligned}
\mathcal{H}_{\textsc{tb}}&\!=\!\sum_{i,j,\alpha,\beta,\sigma}
(t_{i\alpha,j\beta}d^\dagger_{i\alpha\sigma}p_{j\beta\sigma}\!+\!\mathrm{H.c.})
+\sum_{i,\alpha}E_{i\alpha}n_{i\alpha}\\
&+\!\lambda_{\mathrm{Cr}}\sum_i\mathbf L_i\!\cdot\!\mathbf s_i
+\!\lambda_{\mathrm I}\sum_j\mathbf L_j\!\cdot\!\mathbf s_j
-\!U\sum_i\mathbf m_i\!\cdot\!\mathbf s_i,
\end{aligned}
\label{eq:TB_Hamiltonian_main}
\end{equation}
where $i=1,2$ and $j=1,2$ label the two Cr and two I sites, respectively, while $\alpha,\beta,\sigma$ denote the Cr $d$-orbital, I $p$-orbital, and spin indices, respectively. The model includes Slater--Koster hopping, Cr on-site energies, and SOC on both species. The local field $U\mathbf m_i$ mimics the spin-polarizing effects of Coulomb repulsion and Hund's coupling and constrains the Cr spin direction. Here, $\mathbf s$ denotes the electron-spin operator. Taking the iodine $p$ level as the energy reference, we adopt
$E_{xz}=E_{yz}=-1.8$, $E_{xy}=-1.2$, $E_{z^2}=-0.3$, $E_{x^2-y^2}=0.5$, $\lambda_{\rm I}=0.1$, $\lambda_{\rm Cr}=0.005$ and $U=2.5$, all in eV. 
The TB model reproduces the opposite Kitaev signs. The considered cluster contains 20 valence electrons and we thus evaluate the energy by summing the 20 lowest eigenvalues. Its spin-direction dependence is then mapped onto $H=\sum_{\eta}J_\eta S_1^\eta S_2^\eta$, where $\mathbf S_1$ and $\mathbf S_2$ are classical unit spins and $\eta=x,y,z$.  As shown in \textcolor{blue}{Fig. S3} \cite{sm}, the TB model qualitatively reproduces AFM $K_{1\mathrm{S}}>0$, FM $K_{1\mathrm{L}}<0$ and $|K_{1\mathrm{S}}|<|K_{1\mathrm{L}}|$, consistent with DFT. 

We now examine the origin of this sign contrast. With the hopping amplitudes fixed, both Kitaev couplings are FM for degenerate $t_{2g}$ levels. Raising $E_{xy}$ relative to $E_{xz}=E_{yz}$ reverses $K_{1\mathrm{S}}$ to AFM, whereas $K_{1\mathrm{L}}$ remains FM (\textcolor{blue}{Figs. S4 and S5} \cite{sm}). This reversal persists when the $e_g$ levels are kept degenerate. In contrast, splitting the $e_g$ levels while keeping the $t_{2g}$ manifold degenerate has little effect. This seemingly points to $t_{2g}$ splitting as the sole origin. However, the reversal disappears when the $e_g$ sector is excluded from the hopping network. Together, these results indicate that Jahn--Teller-induced $t_{2g}$ splitting and the coupled $t_{2g}$--$e_g$ hopping network induce the sign contrast [see \textcolor{blue}{Sec.~II} of SM~\cite{sm}].

To clarify their microscopic interplay, we analyze ligand-SOC-assisted virtual excitations in a weak-SOC control model with equal Cr-I bond lengths. On the short bond [\figref{fig:guidao}(c)], the $d_{xy}$ orbital couples to $d_{yz}$ and $d_{xz}$ via $L_xs_x$ and $L_ys_y$, respectively, and to the $e_g$ orbitals through ligand $p$ orbitals. Raising $E_{xy}$ changes orbital mixing and excitation energies, rebalancing spin-conserving and spin-flip contributions to the directional FM--AFM energy differences and reversing both transverse Kitaev responses from negative to positive. On the long bond [\figref{fig:guidao}(d)], $L_ys_y$ and $L_zs_z$ instead probe $d_{xy}\leftrightarrow d_{z^2}$ and $d_{xy}\leftrightarrow d_{yz}$ connections. The former response varies weakly, whereas the latter becomes less negative. Both remain negative and outweigh the positive $L_xs_x$ contribution, preserving $K_{1\mathrm{L}}<0$. 
Thus, the sign contrast arises from crystal-field splitting that modifies ligand-SOC-assisted excitations differently on inequivalent bonds, providing a microscopic route to realizing opposite Kitaev signs.

\begin{figure}[!tbp]
  \centering
  \includegraphics[width=8cm]{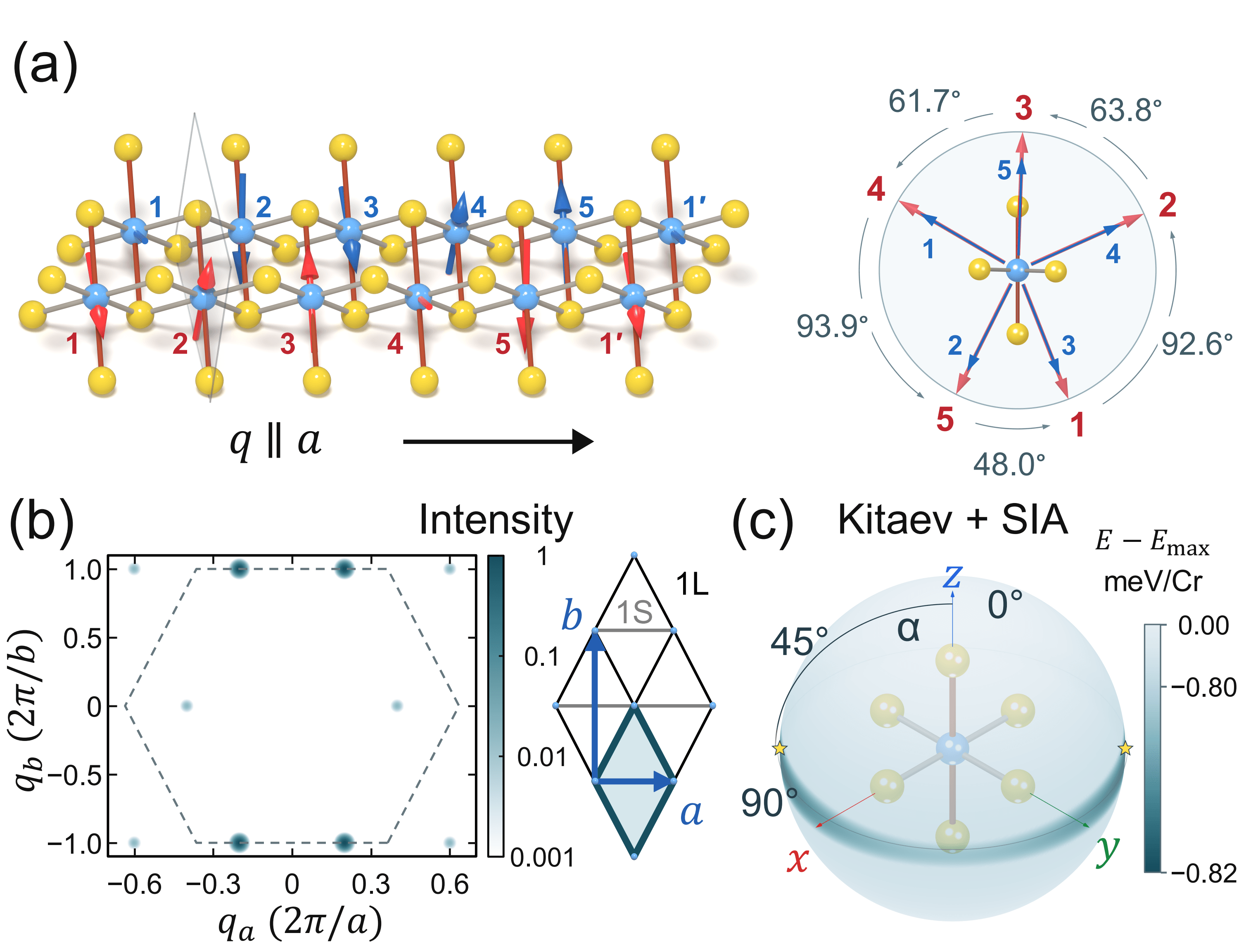}
  \caption{Magnetic ground state and anisotropy of CrI$_2$. (a) Proper-screw ground state propagating along ${\bm a}$ with a period of $5a$. The transparent rectangle indicates the spin-rotation plane.  Projection of the five spins within one period onto this plane reveals nonuniform angular spacing.  (b) Spin structure factor $S(\mathbf q)$ obtained from large supercell simulations, with main peaks at $(q_a,q_b)=(\pm0.2,\pm1)$ and also anharmonic peaks at $(\pm0.4,0)$. The inset shows the Cr lattice and primitive cell. (c) Spherical energy map for Kitaev exchange and single-ion anisotropy (SIA). Each direction on the sphere represents the energy of a spin-rotation plane with its normal along that direction. The CrI$_6$ octahedron inside the sphere provides an orientational reference. }
  \label{fig:gs}
\end{figure}

\begin{table}[bp]
\centering
\caption{The DFT-derived magnetic parameters in \eqnref{eq:heff}. The spins are of unit length. All parameters are in meV.}
\renewcommand{\arraystretch}{1.2}
\begin{tabular*}{\columnwidth}{@{\extracolsep{\fill}} l r l r l r}
\hline\hline
$J_{1\mathrm{S}}$ & 2.85 &
$J_{1\mathrm{L}}$ & 4.32 &
$B_{1\mathrm{L}}$ & $-0.35$ \\

$K_{1\mathrm{S}}$ & 0.48 &
$K_{1\mathrm{L}}$ & $-1.52$ &
$B_{1\mathrm{S}}$ & $-0.82$ \\

$J_{3\mathrm{S}}$ & 1.44 &
$J_{3\mathrm{L}}$ & $-0.97$ &
$A_{n}$ & $-0.29$ \\
\hline\hline
\end{tabular*}
\label{table:coeff}
\end{table}

\textcolor{blue}{Ground state and its anisotropy.}
We now analyze the whole spin Hamiltonian and determine its ground state. As shown in \tabref{table:coeff}, the fitted parameters reveal sizable competing AFM Heisenberg interactions: $J_{1 \rm L}=4.32$ meV is stronger than $J_{1 \rm S}=2.85$ meV; $J_{3 \rm S}=1.44$ meV is also AFM, while $J_{3 \rm L}=-0.97$ meV is FM. Previous first-principles studies have reported substantially different Heisenberg interactions in CrI$_2$, including different signs of the leading exchanges~\cite{schneeloch_helimagnetism_2024,acharya_first-principles_2025,widyandaru_tunable_2025,yu_sliding_2025}. Nevertheless, they consistently indicate that competing interchain and/or longer-range interactions are important~\cite{schneeloch_helimagnetism_2024,widyandaru_tunable_2025,yu_sliding_2025}.
In contrast, although the Heisenberg couplings vary appreciably with Hubbard $U$, strain, and functional, the nearest-neighbor Kitaev interactions robustly retain opposite signs, $K_{1 \rm S}>0$ and $K_{1 \rm L}<0$, across all tested conditions, including HSE [see \textcolor{blue}{Sec.~I(e)} of SM~\cite{sm}]. 
Both $B_{1 \rm S}$ and $B_{1 \rm L}$ are negative and therefore favor collinear spin correlations, whereas the fitted off-diagonal $\Gamma$ interactions are negligibly small.


The competition among these interactions stabilizes a proper-screw ground state propagating along the ${\bm a}$ lattice vector direction (1S bond direction), with the spin-rotation plane perpendicular to the propagation vector [\figref{fig:gs}(a)]. Monte Carlo simulations followed by conjugate-gradient relaxation give a period of $5a$, consistent with the reported period of approximately $5a$ in monolayers \cite{fuyingshuang}. In contrast, the bulk experimental wave vector $\mathbf q=(0.2492,0,0)$ corresponds to a period of approximately $4.01a$ \cite{schneeloch_helimagnetism_2024}. Agreement with the bulk experiment therefore concerns the spin-rotation plane and propagation direction, rather than the modulation period. Moreover, the proper-screw state is found to be strongly anharmonic. As shown in \figref{fig:gs}(a), the angles between neighboring spins range from $48.0^\circ$ to $93.9^\circ$, compared with $72^\circ$ for a uniform five-site spiral. This pronounced  modulation generates higher-harmonic peaks at $(\pm0.4,0)$ in the spin structure factor [\figref{fig:gs}(b)]. Similar helical bunching driven by SIA has been reported, where the inferred turn angles span a range of $45.4^\circ$--$57.4^\circ$~\cite{chaix2016helical}, substantially narrower than that found here.

We finally examine the origin of the proper-screw anisotropy and its anharmonicity. For the mixed-sign Kitaev interactions, the energy remains invariant when the spin-rotation plane is rotated about the $z$ axis. Thus, the Kitaev interaction alone does not distinguish the proper-screw and cycloidal configurations. By contrast, the SIA favors spin alignment close to the ${\bm n}$ axis. The proper-screw anisotropy therefore emerges from the interplay between the mixed-sign Kitaev exchange and SIA [\figref{fig:gs}(c)]. Furthermore, considering SIA and Kitaev anisotropy separately yields turn angle ranges of $57.5^\circ$--$66.1^\circ$ and $48.3^\circ$--$89.9^\circ$, respectively. Similar nonuniform spin rotation was recently reported in CrI$_2$~\cite{schneeloch_helimagnetism_2026}; our results suggest that the anharmonicity here stems mainly from inequivalent Kitaev couplings. By comparison, NiI$_2$, with same-sign Kitaev couplings on different bonds, exhibits uniform spin rotation \cite{li_realistic_2023}, highlighting the role of opposite-sign Kitaev exchange in the pronounced angular modulation found here.

\begin{figure}[!t]
  \centering
  \includegraphics[width=0.7\columnwidth,trim=0 15bp 0 45bp,clip]{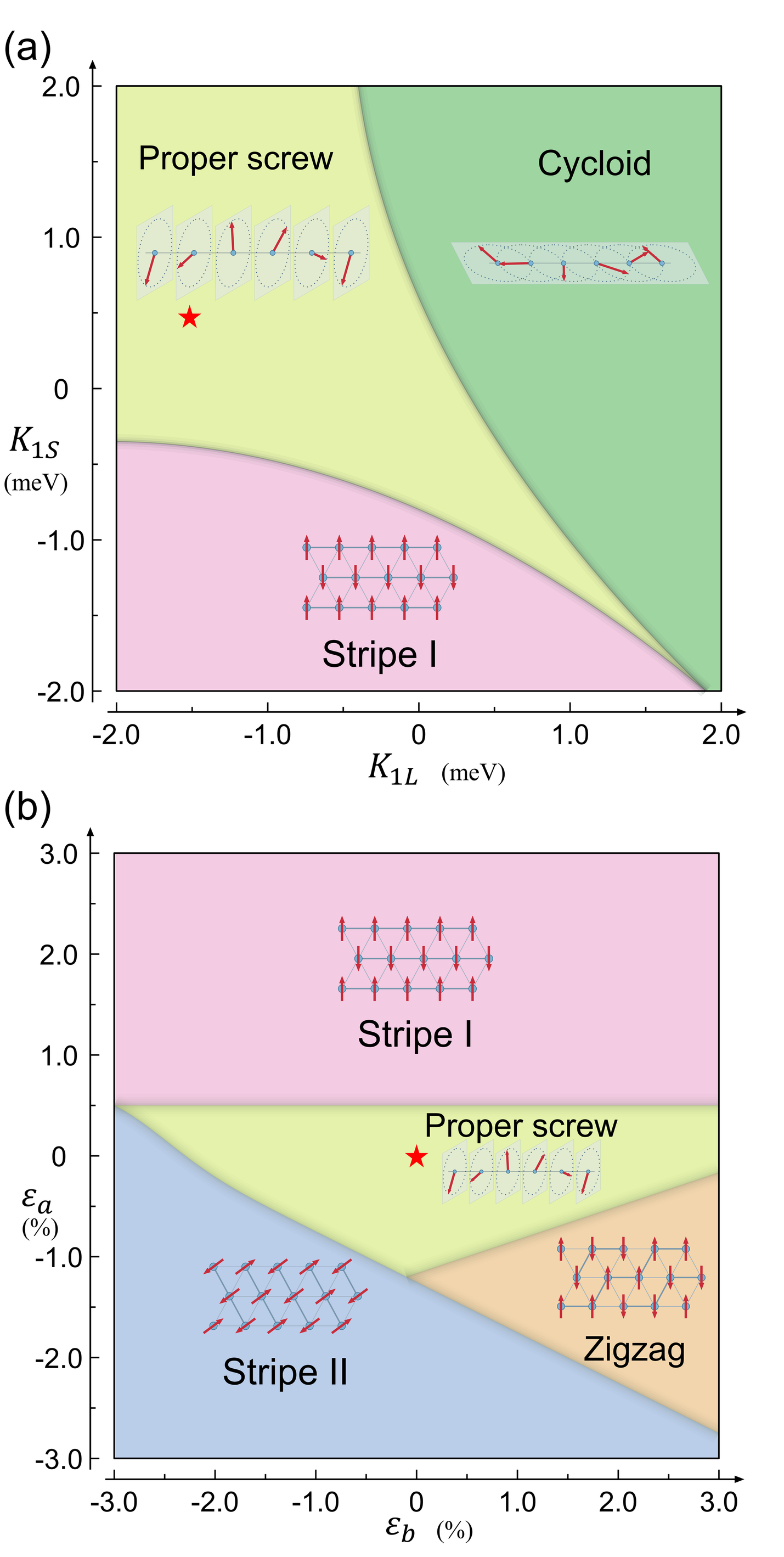}
  \caption{Magnetic phase diagrams as functions of (a) $K_{1\mathrm{L}}$ and $K_{1\mathrm{S}}$, with all other interactions held at their fitted values, and (b) biaxial strain $\varepsilon_a$ and $\varepsilon_b$, with magnetic parameters recalculated and refitted at each strain point. In (a), the colors summarize the three dominant magnetic states and the smooth boundaries guide the eye. Table~S6 of the Supplemental Material lists the matched-magnitude sign controls. Modulated-collinear and complex/multipeak candidates are omitted from the schematic. Stripe I and II denote collinear antiferromagnetic states with ferromagnetic chains along two inequivalent nearest-neighbor bond directions. The zigzag state consists of antiferromagnetically coupled ferromagnetic zigzag chains. Stars mark the reference Kitaev couplings from \tabref{table:coeff} in (a) and zero strain in (b).}
  \label{fig:phases}
\end{figure}

To further clarify the role of Kitaev signs, we vary $K_{1\mathrm L}$ and $K_{1\mathrm S}$ while keeping all other parameters fixed at the values listed in \tabref{table:coeff}. The phase diagram in \figref{fig:phases}(a)  shows a broad proper-screw region for $K_{1\mathrm L}<0$ and $K_{1\mathrm S}>0$. With the opposite sign pattern ($K_{1\mathrm L}>0$ and $K_{1\mathrm S}<0$), this state survives only within a narrow window between the stripe-I and cycloidal phases. Strong AFM $K_{1\mathrm L}$ favors a cycloidal state with its spin-rotation plane parallel to the monolayer. Sizable Kitaev exchange is thus compatible with easy-plane anisotropy, reminiscent of the in-plane cycloid in CoI$_2$~\cite{Kuindersma1981NiI2CoI2,Kim2023CoI2}. In contrast, strong FM $K_{1\mathrm S}$ stabilizes stripe-I AFM order, with FM chains along ${\bm a}$ and spins aligned along $\pm z$.

Strains are applied along ${\bm a}$ and ${\bm b}$, and the spin Hamiltonian is refitted at each strain point, allowing for variation in the measured lattice constants. As summarized in \textcolor{blue}{Sec.~V} of the SM~\cite{sm}, $J_{\rm 1S}$ can change sign, while all other parameters retain theirs. The phase diagram in \figref{fig:phases}(b) shows that proper-screw order persists over a broad range of strain along ${\bm b}$, whereas tension along ${\bm a}$ above approximately $0.5\%$ stabilizes stripe I order. The stripe I order was predicted at zero strain in previous works \cite{nunez-murillo_single-layer_2025,widyandaru_tunable_2025,valdes-toro_stacking-controlled_2025}. The zigzag phase emerges in the lower-right region under combined ${\bm b}$-axis tension and ${\bm a}$-axis compression, with spins aligned along $\pm z$. The stripe II phase occupies the lower-left and bottom regions and has two degenerate variants, with FM chains along two different long bond directions, respectively.
No FM state is observed in the studied strain range.

\textcolor{blue}{Conclusions.} In summary, it is demonstrated that Jahn--Teller splitting of CrI$_2$ can differentiate not only the strengths but also the signs of Kitaev exchange through the bond-dependent response of coupled $t_{2g}$--$e_g$ hopping networks. The resulting mixed-sign couplings cooperate with single-ion anisotropy to select the proper-screw rotation plane, while exchange and strain tuning reveal competition between this state and cycloidal or collinear orders. These findings identify the bondwise sign pattern as an additional variable for understanding and designing  Kitaev magnets. The realization of mixed-sign Kitaev interactions invites exploration of magnetic order, collective excitations, and quantum correlations beyond the uniform-sign settings.

\begin{acknowledgements}
The work at Fudan is supported by National Key R\&D Program of China (Grant No. 2022YFA1402901), NSFC (Grants No. 12188101 and No. 12274082), Quantum Science and Technology-National Science and Technology Major Project (2024ZD0300102), Shanghai Science and Technology Program (No. 23JC1400900), Shanghai Pilot Program for Basic Research-Fudan University (No. TQ1400100) (Shanghai ST Program, Grant No. 23TQ017), Guangdong Major Project of the Basic and Applied Basic Research (2021B0301030005), the robotic AI-Scientist platform of Chinese Academy of Sciences, and New Cornerstone Science Foundation. C.X. also acknowledges Scientific Research Innovation Capability Support Project for Young Faculty of the Ministry of Education (U40) and Xiaomi Young Talents Program.
\end{acknowledgements}

\nocite{schneeloch_helimagnetism_2024,kresse_efficiency_1996,kresse_efficient_1996,blochl_projector_1994,kresse_ultrasoft_1999,perdew_restoring_2008,dudarev_electron_1998,ma_constrained_2015,lou_pasp_2021,pasp_manual_2023,li_constructing_2020,hukushima_exchange_1996,jain_materials_2013,perdew_generalized_1996,heyd_hybrid_2003,krukau_influence_2006,harrison_electronic_1989,slater_simplified_1954}
\bibliography{references}

\end{document}